\def\bbibitem#1{\bibitem{#1}}

\documentstyle[prb,aps,amsfonts,multicol]{revtex}

\newcommand{\g}{\sigma}
\newcommand{\is}{_{i,\g} }
\newcommand{\beq}{\begin{equation}}
\newcommand{\eeq}{\end{equation}}

\newcommand{\up}{\uparrow}
\newcommand{\down}{\downarrow}
\newcommand{\al}{\alpha}

\newcommand{\de}{\partial}
\newcommand{\tr}{\hbox{\rm tr}\,}

\ifx\eqnarraynn\undefined
\newcommand{\beqn}{\begin{eqnarray}}
\newcommand{\eeqn}{\end{eqnarray}}
\else
\newcommand{\beqn}{\begin{eqnarraynn}}
\newcommand{\eeqn}{\end{eqnarraynn}}
\fi
\newcommand{\n}[2]{#1_{#2}^{\dag} #1_{#2}}
\newcommand{\eqref}[1]{(\ref{#1})}

  \def\btwocols{}
  \def\etwocols{}

\def\makecols{
  \def\btwocols{\begin{multicols}{2}}
  \def\etwocols{\end{multicols}}
}

\ifpreprintsty\else\makecols\fi

\def\evi{\it}

\newcommand{\dw}{\down}
\newcommand{\vecs}[1]{ |#1\!>_N\; } 
\newcommand{\vecr}[1]{ |#1)_N\; } 
\newcommand{\vecrt}[1]{ \;_N(#1| } 
\newcommand{\NU}{N_{\up}}
\newcommand{\ND}{N_{\dw}}

\begin{document}


\draft   

\title{
On an exact criterion for
 choosing the hopping operator
\\
in the
four-slave-boson approach
}

\author{E. Arrigoni
\cite{pa1}
}
\address{
Max-Planck-Institut f\"ur Physik
Complexer Systeme, (Aussenstelle) D-70569 Stuttgart , Fed. Rep. Germany }

\author{ G.C. Strinati
\cite{pa2}}
\address{
 Dipartimento di Fisica, III Universit\`a di Roma,
 I-00146 Roma, Italy
}

\maketitle

\begin{abstract}
We consider the $N$-component generalization of the
four-slave-boson approach to the Hubbard model, where  $1/N$ acts as the small
parameter  that  controls the fluctuations  about the saddle point,
and address the problem of the appropriate choice of the bosonic
hopping operator $z_{i\g}$.
By suitably reorganizing the Fock space,
we show that the square-root form for $z\is$ (originally
introduced by Kotliar and Ruckenstein) reproduces the exact
independent-fermion  ($U=0$) results not only at the mean-field
($N=\infty$)
level  but also to  {\evi all orders in the
$1/N$ expansion}, provided one relaxes the
 usually adopted normal-ordered prescription for
$z\is$.
This  ensures that  $z\is$  needs not  be
modified  at successive orders in the fluctuation
expansion,  and implies that all correlation
functions are correctly recovered in the $U=0$
limit, a nontrivial result for the slave-boson approach.
 In addition,
it  provides a
stringent requirement on the form of $z\is$, which might be also
generalized to alternative slave-boson
formalisms (like the spin-rotation-invariant formulation).
\end{abstract}

\pacs{ PACS numbers :  71.27.+a, 03.65.-w, 71.10.+x }


\btwocols

The four-slave-boson method  has been widely used in the past few
years to deal with  the Hubbard model (with  on-site repulsion
$U$). This method is usually implemented  via a mixed fermion-boson
functional integral, which allows for a systematic expansion in terms
of fluctuations about
an appropriate mean field.
The fluctuation expansion is, in turn, controlled by introducing an
additional fermion degeneracy   $N$ and using $1/N$ as the
expansion parameter.\cite{low+cli}
Already at the mean-field level,
the method provides a reasonable description of the phase diagram
and of several static quantities for the Hubbard model,
 as  shown by comparisons with Quantum Monte
Carlo calculations.\cite{qmc}
Since the slave-boson approach should be, by its conceiving,
appropriate to the strong-coupling regime, it
  might  not be expected  to reproduce
  the noninteracting fermion
limit at the same time.
In this respect, Kotliar and Ruckenstein (KR) proposed   choosing
the
 hopping operator $z_{i,\g}$ (which, due to the
redundancy of the slave-boson Fock space, is not uniquely defined)
  to obtain  the
correct result  at the mean-field level
in the $U\to 0$ limit.\cite{kr}
It has, however,  later been questioned whether the same form of $z_{i,\g}$
 remains appropriate when  corrections beyond mean field
 are introduced.\cite{legu} It turns out, in fact, that unphysical
 results occur  when
the KR
form for $z\is$ is used beyond  mean field.\cite{legu,prl}

It has been  recently shown
that the above anomalies stem from an
inappropriate  operator ordering
of $z_{i,\g}$.\cite{low+cli}
In fact,  previous to the work of Ref. \onlinecite{low+cli},
all slave-boson calculations have been invariably carried out by  assuming the
normal-ordering prescription for
$z_{i,\g}$,
since  normal ordering allows
 for a direct functional-integral
representation of the Hamiltonian.
 If one adopts, instead, the KR form of
 $z_{i,\g}$ {\evi without}
modifying the original operator ordering,
 fluctuation corrections
do not  spoil the  $U=0$
mean-field results (at least, for the
free energy and related static quantities),
in contrast to the normal-ordered version for $z_{i,\g}$.
The question naturally arises whether the results of Ref.
\onlinecite{low+cli}, which have been obtained
up to leading
 order beyond mean field,
  also hold when higher-order
 corrections are considered, and not only for static
  but also for dynamic quantities.

In this paper,  by exploiting  some exact properties of the
hopping operator and of the related
constraints for arbitrary values of the
degeneracy parameter $N$, we show
 that  the $U=0$ independent-fermion result can be recovered
(at zero temperature)
to {\evi all orders} in the $1/N$ expansion,  {\evi provided}
the
 strict square-root
  form for $z_{i\g}$  is used
 instead of its normal-ordered version.
This result holds
not  only for
the ground-state energy, but also for
the ground-state expectation values of
operators written in terms of the {\evi physical} fermion operators
(obtained as
 the product of $z_{i\g}$ with a pseudofermion operator).

We consider the $N$-component generalization of the
slave-boson single-band Hubbard Hamiltonian, \cite{low+cli,kr} of the form
\beqn
\label{ham} 
 H  && =  \sum_{i,i'} \sum_{\g=\pm 1}\sum_{S=1}^N  t_{i,i'}
f^{\dag}_{i,S,\g} z^{\dag}\is
 z_{i',\g} f_{i',S,\g}
\nonumber \\&&
+  U \sum_i \n di
\eeqn
supplemented by
the constraints
\begin{mathletters}
\beqn
\label{consa}
&& \n di + \sum_{\g=\pm1} \n s{i,\g} + \n ei  =N \;,
\\&&
\label{consb}
\sum_{S=1}^N \n f{i,S,\g} = \n s{i,\g} +
 \n di   \;
\eeqn
\label{cons}
\end{mathletters}\noindent
which are  straightforward generalizations of the $N=1$ case
considered by KR. Although
for $N>1$ the constraints \eqref{cons} no longer guarantee
a one-to-one correspondence between the original fermion and
the fermion-boson problems, we consider
 this  not an issue since  $1/N$ has been
here introduced merely as an
expansion parameter and the   $N=1$ value will eventually be selected.
In \eqref{ham} and \eqref{cons}
$e_i,s_{i,\g}$, and $d_i$ are slave-boson operators
 associated with  empty, singly, and doubly occupied states, in the order,
$f_{i,S,\g}$ is a pseudofermion operator  with spin $\g$ and
component $S$, and  $i$ ($i'$)  extends over all
 lattice sites.
For arbitrary $N$,
the presence of the bosonic operator $z\is$ in \eqref{ham} is
required  to leave  the
 subspace identified by the constraints \eqref{cons} invariant; its choice,
however,  is to some extent
arbitrary due to the redundancy of the Fock space.
Kotliar and Ruckenstein  exploited this arbitrariness and
suggested using the  form\cite{kr}
\beq
\label{zr} 
 z_{i,\g}=  s^{\dag}_{i,-\g} R_{i,\g} d_i + e^{\dag}_i R_{i,\g} s_{i,\g} \;,
\eeq
with
\label{rkrsq} 
\begin{mathletters}
\beq
 R_{i,\g} =  R_{i,\g}^{KR} =  :R_{i,\g}^{SQ}:
\label{rkr} %
\eeq
and
\beqn
  R_{i,\g}^{SQ} && = \frac{1}{\sqrt{N- \n di - \n s{i,\g}}}
\nonumber\\&&\times  
\frac{1}{\sqrt{N- \n ei - \n s{i,-\g}}} \;,
\label{rsq} %
\eeqn
\end{mathletters}\noindent
in order to  reproduce the exact $U=0$ result at the mean-field
level.
The normal ordering $:R_{i,\g}^{SQ}:$ in \eqref{rkr} stems by requiring
a  simple mapping of the Hamiltonian \eqref{ham} into the corresponding
Action of the functional integral.

 It has, however, been shown
in Ref. \onlinecite{low+cli} that
improved results are obtained beyond mean field by relaxing the normal
ordering in Eq. \eqref{rkr} (whereby the normal-ordering requirements
of the
functional integral are met by suitably reordering the operators in
\eqref{rsq} within the $1/N$ expansion).
In the following, we shall   not rely on the functional-integral
 formulation; rather, we will prove exact results
 {\evi for arbitrary
values of $N$}  using the operator form \eqref{rsq}.
Specifically, we will show that:
(i) the mixed fermion-boson Fock space restricted
by \eqref{cons}  can be split into subspaces (labeled by a quantum
number $J_i$) that remain invariant under
the action of
 $z\is$ when  the form \eqref{rsq}  is adopted;
(ii) a particular subspace  can be
identified (with $J_i=0$ for all $i$),  where
a one-to-one
correspondence between matrix elements of
the original fermion and  of the fermion-boson
Hamiltonians
can be established for all $N$.
We will also argue that, when $U=0$, the ground state belongs to the
 subspace $\{J_i=0\}$.

We begin by analizing the properties of the  operator \eqref{rsq} in the
 bosonic Fock space at  a given site $i$.
We specify
 a generic basis state in this space
 via the bosonic occupation numbers $ n_e, n_{\up}, n_{\dw}$, and
$n_d$ associated with
 $\n e{}$, $\n s{\up}$
$\n s{\dw}$, and $\n d{}$, in the order:
\beq
 |n_e, n_{\up},n_{\dw}, n_d> \;.
\label{bstate}
\eeq
 The  operator  \eqref{rsq}
 is diagonal
in the representation \eqref{bstate}
(by contrast, its normal-ordered version \eqref{rkr} has
nontrivial matrix elements in this representation).
This allows us to verify the following
commutation relations for the operator $z^{SQ}_{i,\g}$  given by \eqref{zr}
with $R_{i,\g}=R_{i,\g}^{SQ}$, when acting on any state of the form
\eqref{bstate}
(we will omit below, for simplicity, the site index $i$ and the label
$SQ$):
\beq
[z_{\up},z_{\dw}]=0 \quad,\quad [z_{\up},z^{\dag}_{\dw}]=0 \;,
\label{comm}
\eeq
plus their Hermitian conjugates.

The  ``physical'' (fermion-boson)
subspace is identified  by the
constraints \eqref{cons}.
The set of bosonic states associated with a given
 fermionic configuration is thus determined via  the pair of
 fermionic occupation numbers
$N\is=\sum_{S=1}^N \n f{i,S,\g}$ ($\g=\pm1$) [a given pair ($N_{\up}$,
$N_{\dw}$) may, on the other hand, be
   associated with more than one fermionic configurations].
The constraints
\eqref{cons} thus associate the pair
($N_{\up},N_{\dw}$) with
the bosonic subspace
spanned by the basis states
\beqn
\vecs{N_{\up}&&,N_{\dw}; n_d}
\nonumber \\ &&
=|N-N_{\up}-N_{\dw}+n_d, N_{\up}-n_d, N_{\dw}-n_d , n_d>
\label{classal}
\eeqn
where $n_d$ can take
$1+min(N_{\up},N_{\dw},N-N_{\up},N-N_{\dw})$ integer values ranging
within
 $max(0,-N+N_{\up}+N_{\dw}) \leq n_d \leq  min(N_{\up},N_{\dw})$.
The  subspaces
$\{\vecs{N_{\up},N_{\dw}; n_d}\}$
with different values of $(N_{\up},N_{\dw})$ are
 connected by the
 operators $z_{\g}$ and $z^{\dag}_{\g}$.
Consider, in particular, the application of the
   operators  $z^{\pm}_{\g}$ (with $z^{+}_{\g}=z^{\dag}_{\g}$ and
$z^{-}_{\g}=z_{\g}$)
on the  state
$\vecs{n,0;n_d=0}$:
\beq
z^{\pm}_{\up} \vecs{n,0;n_d=0}=C^{\pm}_n \vecs{n\pm 1,0;n_d=0}\;,
\label{zdup}
\eeq
 where the normalization constants
 $C^{\pm}_n$  equal  unity
only
for the $SQ$ form
\eqref{rsq}.\cite{lastel}
Therefore,
the operators  $z^{\pm}_{\up}$  act as creation and
destruction operators in the subspace spanned by
$ \{ \vecs{n,0;n_d=0};  n=0,\cdots,N \} $.
The above conclusions hold, as well, if one exchanges up and down spins.
Exploiting  the commutation relations \eqref{comm}, one can then
unambiguously  define
the subspace  of the physical space spanned by the $(N+1)^2$ (normalized)
states
\beq
\vecr{J=0;N_{\up},N_{\dw}}
 \equiv
(z^{\dag}_{\up})^{N_{\up}}
(z^{\dag}_{\dw})^{N_{\dw}}  \vecs{0,0;n_d=0}
\label{veck}
\eeq
with $0\leq (N_{\up},N_{\dw})\leq N$.
Here $J$ is a new {\evi quantum number} which will be essential for the
following arguments.

Consider, next, the
 subspace with $N_{\up}=N_{\dw}=1$
 spanned by the two
states
$\vecs{1,1;n_d=0}$ and $\vecs{1,1;n_d=1}$ (cf. Eq. \eqref{classal}),
 since $n_d$  takes the values $0$
and $1$ in this case.
Recall  that the state
$\vecr{J=0;1,1}= z^{\dag}_{\up} z^{\dag}_{\dw}  \vecs{1,1;n_d=0}$
belongs to this subspace.
The complement of this
state in the subspace we are considering
can then be found by Schwartz orthogonalization.
 Let us denote
this state by $\vecr{J=1;1,1}$.
Similarly to what we have done in Eq. \eqref{veck}, from the state
$\vecr{J=1;1,1}$ one can construct a whole set of $(N-1)^2$ companion
states as follows:
\beqn
 C_{(N_{\up},N_{\dw})}^{(J=1)} && \vecr{J=1;N_{\up},N_{\dw}}
\nonumber \\ &&
 = (z^{\dag}_{\up})^{N_{\up}-1}  (z^{\dag}_{\dw})^{N_{\dw}-1}
\vecr{J=1;1,1} \;,
\label{veck1}
\eeqn
where
 $C_{(N_{\up},N_{\dw})}^{(J=1)} $ is a
 normalization constant which is no longer unity as it was
in Eq. \eqref{veck}.
This difference between the values of the normalization constants for
the cases $J=0$ and $J\ge1$ is due to the square-root choice
\eqref{rsq} and will be important in the following.
It can further be shown that the states of the two sets
$\{ \vecr{J=0;\NU,\ND} \}$
and $\{ \vecr{J=1;\NU,\ND} \}$ are mutually orthogonal.

One can then proceed by induction and
 identify $[N/2]+1$ orthogonal subspaces,
labeled by the quantum number $J(=0,1,\cdots,[N/2])$ and spanned by the
$(N-2 J+1)^2$ states
\beq
  C_{(N_{\up},N_{\dw})}^{(J)} \vecr{J;N_{\up},N_{\dw}}=
 (z^{\dag}_{\up})^{N_{\up}-J}  (z^{\dag}_{\dw})^{N_{\dw}-J}
\vecr{J;J,J} \;
\label{veckJ}
\eeq
with $J\le (\NU,\ND)\le N-J$.
Note that successive applications of $z_{\g}^{\pm}$ do not change the
value of $J$.

There remains to evaluate
 the constants $C^{(J)}_{(N_{\up},N_{\dw})}$
($=C^{(J)}_{(N_{\dw},N_{\up})}$,
 by symmetry).
One can readily show  that
the quantity
\beq
\al^{(J)}_{n}\equiv
\left| \frac{C^{(J)}_{(n+1,m)}}{C^{(J)}_{(n,m)}} \right|^2 =
\vecrt{J;n,m} \;
  z_{\up} z^{\dag}_{\up} \; \vecr{J;n,m}
\label{cvsal}
\eeq
is eigenvalue of the operator
$ z_{\up} z^{\dag}_{\up}$ associated with the
eigenstate $\vecr{J;n,m}$ and is
 independent of $m$.
 Consider then the trace of the operator $z_{\up} z^{\dag}_{\up}$
in the subspace with given values of $\NU=n$ and $\ND=m$, where, for the sake
of definiteness, we choose $m\le n\le N-m$. This subspace is spanned
by $m+1$ states, which we can specify alternatively by
$\{\vecr{J;n,m}$, with $0\le J \le m \}$
or by $\{\vecs{n,m;n_d}$, with $0\le n_d \le m \}$.
In the first
(diagonal) basis the {\evi restricted} trace takes the form
\beq
\tr(z_{\up} z^{\dag}_{\up})_{(n,m)}  = \sum_{J=0}^m \al_n^{(J)} \;,
\label{tr}
\eeq
from which
the eigenvalue $\al_n^{(m)}$ can be  obtained
as the difference
\beq
\al_n^{(m)} = \tr(z_{\up} z^{\dag}_{\up})_{(n,m)}
- \tr(z_{\up} z^{\dag}_{\up})_{(n,m-1)} \;.
\label{expal}
\eeq
In the second (nondiagonal) basis, on the other hand, the
restricted trace can be readily evaluated since the operator
$z_{\up} z^{\dag}_{\up}$ admits a simple representation.
 Comparing the two results, one  obtains:
\beq
\al_n^{(J)} =
 \frac{(n+1-J)(N-n-J)}{(N-n)(n+1)} \;.
\label{valal}
\eeq
Note that $\al_n^{(0)}=1$, as anticipated, and that  $\al_n^{(J)}=0$
for the upper value $n=N-J$
(except for $J=0$, see
footnote \onlinecite{lastel}).  In addition, we have
\beq
0 \leq \al_n^{(J+1)} < \al_n^{(J)} < \al_{n'}^{(J=0)} =1 \;
\label{prop}
\eeq
for any pair $(n,n')$ and $J\ge1$.

In conclusion, we have shown that
(at any given lattice site)
 the constrained bosonic Fock space
(for given $N$) can be split
 into the direct sum of subspaces
 spanned by the states
$\{\vecr{J;\NU,\ND}$, with $J=0,\cdots,[N/2]\}$.
Within each $J$ subspace,
the operators $z_{\g}$ and $z^{\dag}_{\g}$ act as ``lowering'' and
``raising'' operators with normalization constants given by
\eqref{cvsal} and \eqref{valal} ($C_{(J,J)}^{(J)}=1$, by definition)
and  destroy the extremal states with $N_{\g}=J$ and $N_{\g}=N-J$,
respectively.
In this respect, the $J$ subspaces bear some analogy with the sets of
states obtained in elementary quantum mechanics by coupling two
angular momenta.

We  now return to the (physical) mixed fermion-boson Fock space.
The constraints \eqref{cons} associate
 every fermionic configuration with given values of
$N_{\up}$ and $N_{\dw}$, with the whole set of bosonic states
$\{\vecr{J;\NU,\ND};J=0,\cdots,\min(N_{\up},N_{\dw},N-N_{\up},N-N_{\dw})\}$
(and not
 just with a single bosonic state, like for $N=1$). This implies that
a generic basis
state of the mixed  Fock space can be written (at any given lattice
site) as the product of a
fermionic  and of a bosonic state, as follows:
\beq
 |\phi(N_{\up},N_{\dw})\!> \;
  \vecr{J;N_{\up},N_{\dw}}
\label{phinmk}
\eeq
where $\phi(N_{\up},N_{\dw})$ is a generic fermionic configuration consistent
with the pair $(\NU,\ND)$. Note
that the ``physical'' fermion
operator $z_{\g} f_{S,\g}$
does not change the quantum number $J$ when applied to \eqref{phinmk}.

Concerning further the
 Fock space for the whole lattice, this can also be split into distinct
subspaces,  each identified by a set of integers $\{J_i\}$
with $i$ ranging over all lattice sites.
 Each subspace specified by the set $\{J_i\}$ is invariant under the
slave-boson Hamiltonian with $U=0$, since $\n di$ is the {\evi only} operator
of the Hamiltonian \eqref{ham} that modifies the quantum numbers $J_i$.
 In particular, within the subspace  $\{J_i=0\}$ for {\evi all} $i$
  the operators $z_{i,\g}$ and $z^{\dag}_{i,\g}$  connect states of
  the type \eqref{phinmk}
  with different $\NU$ and $\ND$, without changing their norm. In
  this subspace,   the $U=0$ slave-boson
 Hamiltonian  is thus {\evi
equivalent to its purely fermionic counterpart also
   when $N>1$}.\cite{kr}
This finding is particularly relevant if the
 ground state of the Hamiltonian belongs to the subspace
$\{J_i=0\}$.
In this case,  the
ground-state expectation value of any operator
 in the purely fermionic representation
  equals  the ground-state expectation value of the
corresponding operator
written in the slave-boson
representation, {\evi
provided}  the
operator itself leaves the   subspace  $\{J_i=0\}$ invariant.
This holds, e. g., for the correlation (Green's)
 functions, when the ``physical''
fermion operator $  z_{i,\g} f_{i,S,\g}$ with $R_{i,\g}$ given
by \eqref{rsq}
is used.
Since all physical quantities in the purely fermionic representation
scale simply with integer powers of $N$ when $U=0$, the above
conclusion implies that in the mixed fermion-boson representation all
corrections (over and above the mean-field result) vanish identically
{\evi at any order in $1/N$}.
This result
 confirms and extends on general ground the numerical
results obtained in Ref. \onlinecite{low+cli} at the leading order in $1/N$.

There remains to prove that the ground state of \eqref{ham} with $U=0$
  belongs to the subspace $\{J_i=0\}$,
that is,
\beq
 E_0[\{J_i=0\}] <  E_0[\{J'_i\}] \quad \quad {\rm with} \,\,
\{J'_i\}\not=\{J_i=0\} \;,
\label{gsk}
\eeq
where $E_0[\{J_i\}]$ stands for the lowest eigenvalue in the subspace
specified by the set $\{J_i\}$.
A property similar to \eqref{gsk} can be readily proved for a
single-particle Hamiltonian with
site-dependent hopping, of the form
\beq
H[\{Z\is\}]  = \sum_{i,i'} \sum_{\g=\pm 1}\sum_{S=1}^N t_{i,i'}
 f^{\dag}_{i,S,\g} Z\is  Z_{i',\g} f_{i',S,\g}
\label{hamt}
\eeq
where $t_{ii}=0$ (by assumption) and
 $Z\is$ are real and  positive numbers.
Let  $\tilde E_0[\{Z\is\}]$ be the ground-state energy associated with a given
configuration $\{Z\is\}$.
Then
\beq
\tilde E_0[\{Z\is\}] \leq \tilde E_0[\{Z'\is\}] \;
\label{eZ}
\eeq
whenever
\beq
Z\is \geq Z'\is \geq 0
\label{ZZ}
\eeq
for all $i$ and $\g$, which  can be shown by realizing that
the ground-state expectation value
\beq
< \!\frac{\de H[\{Z\}]}{\de Z\is} \!> \; = \frac {\de
 \tilde E_0[\{Z\}]}{\de Z\is}
\label{dee}
\eeq
is nonpositive for any chosen $i$ and $\g$.
At this point we note that the slave-boson Hamiltonian \eqref{ham}
with $U=0$ can be mapped, {\evi within each subspace} $\{J_i\}$, onto
the purely fermionic Hamiltonian \eqref{hamt}
by replacing  the c-number $Z\is$
with the Hermitian operator $(\al_{N\is}^{(J_i)})^{1/2}$
[obtained by entering the fermionic operator $N\is$ in the place of $n$
in Eq. \eqref{valal}].\cite{by}
Property \eqref{gsk} would thus
 follow from Eqs. \eqref{prop}, \eqref{eZ}, and \eqref{ZZ}
if the
 $\al_{N\is}^{(J_i)}$
could be
 replaced by  c-numbers.
In particular, this is approximately correct
for large $N$ where one can approximate
\beqn
 \al_{N\is}^{(J_i)} &&= 1 - \frac{1}{\bar n\is (1-\bar n\is)}
\frac{J_i}N +
\frac{J}{N^2} \frac{ (1- \bar  n\is + J \bar n\is)}{(1-\bar n\is) \bar n\is^2}
\nonumber\\&& \!\!\!\!\!
+ \frac{(1- 2 \bar n\is)}{\bar n\is^2 (1-\bar n\is)^2} \frac{J_i (N\is -
<\!N\is\!>)}{N^2} +O(N^{-3})
\eeqn
where $<\!N\is\!>= N \bar n\is$ is the self-consistent ground-state value of
$N\is$ for given $\{J_i\}$.
In this way Eq.\eqref{gsk} is validated for a
sufficiently large (albeit finite) value of $N$. In this sense, we have
proved order by order in $1/N$ that the slave-boson representation
does not modify the exact $(U=0)$ independent-fermion
results.\cite{argument}

In conclusion, motivated by the encouraging numerical results obtained
previously
by implementing  the $1/N$ expansion for the four-slave-boson
method correctly,\cite{low+cli} in this paper we have analyzed
 in detail the structure of the
enlarged slave-boson Fock space for $N>1$ and identified an intrinsic
{\evi dynamical symmetry} associated with a novel quantum number
($J$). We have thus been able to tune the four-slave-boson method at
$U=0$ (where this dynamical symmetry holds exactly) by  comparison
with the independent-fermion results.
Our finding considerably limits (and possibly eliminates) the
arbitrariness on the choice of the hopping operator $z\is$.

\def\prl{Phys. Rev. Lett. }
\def\prb{Phys. Rev. B }
\def\ut#1{{\bf #1}}

\etwocols


\begin{references}

\bibitem[*]{pa1}
 Present address:
Institut f\"ur Theoretische Physik, Universit\"at W\"urzburg, D-97074
W\"urzburg, Germany.

\bibitem[**]{pa2}
Present address:
Dipartimento di Matematica e Fisica, Universit\`a di Camerino, I-62032
Camerino, Italy.



\bbibitem{low+cli}
E. Arrigoni and G. C. Strinati, J. Low Temp. Phys.
\ut{99}, 599 (1995); E. Arrigoni and G. C. Strinati, \prb \ut{52},
XXXX (1995).



\bbibitem{qmc}
 L. Lilly,  A. Muramatsu,  and W. Hanke,  \prl
 \ut{65},  1379 (1990).



\bbibitem{kr}
  G. Kotliar and A.E. Ruckenstein  [\prl
\ut{57}, 1362 (1986)] considered only  extrapolation to the
physical case $N=1$, for
which the original fermion and the slave-boson Hamiltonians are, by
definition,
 equivalent.

\bbibitem{legu}
Th. Jolicoeur and J. C. Le Guillou,  \prb \ut{44},
2403  (1991).

\bbibitem{prl}
E. Arrigoni and G. C. Strinati, \prl \ut{71}, 3178
(1993).

\bbibitem{lastel} Problems occurring for the extremal states with
$n=0$ and $n=N$ can be suitably overcome by letting
 the pseudofermion operator act {\evi before} $z^{\pm}_{\g}$
 on any given state.

\bbibitem{by} By our convention, $\al_n^{(J)}=0$ for $n<N-J$ or $n<J-1$.

\bbibitem{argument} An argument to exclude the possible occurrence of
level crossing at finite $N$ [whereby Eq. \eqref{gsk}
 would be
invalidated]
follows from  the last inequality in Eq. \eqref{prop}.

\end{references}
\end{document}